\documentclass[proof]{pasj01}
\usepackage{comment}

\draft

\begin{document}
\Received{}
\Accepted{}

\title{Hermite integrator for high-order mesh-free schemes}

\author{Satoko \textsc{Yamamoto}\altaffilmark{1,2,3}%
\thanks{Example: Present Address is xxxxxxxxxx}}
\altaffiltext{1}{Department of Earth \& Planetary Sciences, Tokyo Institute of Technology, Ookayama, Meguro-ku, Tokyo 152-8550, Japan}
\altaffiltext{2}{RIKEN Advanced Institute for Computational Science, Minatojima-minamimachi, Chuo-ku, Kobe, Hyogo 650-0047, Japan}
\altaffiltext{3}{Department of Planetology, Kobe University, Rokkodaicho, Nada-ku, Kobe, Hyogo 650-0013, Japan}
\email{yamamoto.s.an@geo.titech.ac.jp}

\author{Junichiro \textsc{Makino},\altaffilmark{3,2,1}}



\maketitle

\begin{abstract}
In most of mesh-free methods, the calculation of interactions between
sample points or ``particles'' is the most time consuming. 
When we use mesh-free methods with high spatial orders, the order of the
time integration should also be high.
If we use usual Runge-Kutta schemes, we need to perform the interaction
calculation multiple times per one time step.
One way to reduce the number of interaction calculations is to use Hermite
schemes, which use the time derivatives of the right hand side of
differential equations, since Hermite schemes require smaller number of
interaction calculations than RK schemes do to achieve
the same order.
In this paper, we construct a Hermite scheme for a mesh-free method
with high spatial orders.
We performed several numerical tests with fourth-order Hermite schemes and
Runge-Kutta schemes.
We found that, for both of Hermite and Runge-Kutta schemes, the overall
error is determined by the error of spatial derivatives, for timesteps
smaller than the stability limit. The calculation cost at the timestep size
of the stability limit is smaller for Hermite schemes.
Therefore, we conclude that Hermite schemes are more efficient than
Runge-Kutta schemes and thus useful for high-order mesh-free methods
for Lagrangian Hydrodynamics.
\end{abstract}

\section{Introduction}
\label{sec:intro}
Lagrangian mesh-free methods, in which particles move following the motion
of fluid, have been widely used for astrophysical hydrodynamical
simulations.
In most of mesh-free methods, the calculation of interactions between
particles is the most time consuming part.
Typically, one particle interacts with $\sim$100 neighbor particles, and
thus the cost of interaction calculations dominates the total calculation
cost.
One way to reduce the number of interaction calculations is to use
Hermite schemes, which use the time derivatives of the right hand side of
differential equations, since Hermite schemes require smaller number of
interaction calculations than RK schemes do to achieve the
same order.

In the field of stellar dynamics, the fourth order Hermite scheme
(\cite{1991ApJ...369..200M}; \cite{1992PASJ...44..141M})
is widely used for high-order integration. The basic idea of the Hermite scheme
is to calculate the time derivative of gravitational acceleration directly,
and use it to construct high-order interpolation polynomial.
If we calculate up to $p$-th order time derivative directly, we can achieve
the order of $s(p+1)$ when we use $s$-step linear multistep method, and
in the case of $s=2$, we can achieve the order of $2(p+1)$.
The two-step linear multistep method can be formulated so that it requires
only one force evaluation per timestep.
In the case of grid-based scheme for hydrodynamics, \citet{Aoki1997IDO}
described the method based purely on the Taylor expansion, which achieves
order $p+1$.

In this paper, we combine the Hermite scheme with Consistent Particle
Hydrodynamics in Strong Form (CPHSF; \cite{2017arXiv170105316Y}) which is
one of high-order mesh-free methods.
One disadvantage of the Hermite scheme is that, even though it requires
smaller number of interaction calculation, the calculation cost of one
interaction is higher because we need to calculate high-order derivations.
In the case of CPHSF or other MLS-based interpolation, high-order
interpolation polynomial gives spatial derivatives, and we only need to
convert spatial derivatives to time derivations using the original
differential equations. Thus, the increase of the calculation cost is
small and independent of the number of neighbors.

We performed several numerical tests. Fourth-order Hermite schemes and
second- and fourth-order Runge-Kutta schemes are used for the test with a
periodic boundary, and an implicit Hermite scheme, an implicit
fourth-order Runge-Kutta scheme and the backward-Euler scheme are used for
the test with boundary conditions.
We found that, for both of Hermite and Runge-Kutta schemes, the overall
error is determined by the error of spatial derivatives, for timesteps
smaller than the stability limit. The calculation cost at the timestep size
of the stability limit is smaller for Hermite schemes.
Therefore, we conclude that Hermite schemes are more efficient than
Runge-Kutta schemes and thus useful for high-order mesh-free methods
for Lagrangian Hydrodynamics.

In the rest of this paper, we first present the formulation of the Hermite
scheme for CPHSF in
section \ref{sec:form}, and report the results of numerical tests in
section \ref{sec:test}. We summarize our study in section \ref{sec:discandsum}.

\section{Derivation of the high-order scheme}
\label{sec:form}
In this section, we present the derivation of the fourth-order Hermite
schemes for CPHSF.

\subsection{Hermite scheme}
In this section we present the formulation of the fourth-order Hermite
schemes (\cite{1991ApJ...369..200M}; \cite{1992PASJ...44..141M}).
Consider a second-order differential equation,
\begin{eqnarray}
	\frac{d\boldsymbol{x}}{dt} &=& \boldsymbol{v}, \\
	\frac{d\boldsymbol{v}}{dt} &=& \boldsymbol{a}(\boldsymbol{x}).
\end{eqnarray}
Here, $\boldsymbol{x}$ and $\boldsymbol{v}$ denote the position and
velocity of one particle.
The fourth-order Hermite scheme is derived as follows. The predictor at
time $t_n$ is given by
\begin{eqnarray}
	\label{eq:x,predictor}
	\boldsymbol{x}_p &=& \boldsymbol{x}_n + \boldsymbol{v}_n \Delta t
	+ \frac{\boldsymbol{a}_n}{2} \Delta t^2 + \frac{\boldsymbol{j}_n}{6} \Delta t^3, \\
	\label{eq:v,predictor}
	\boldsymbol{v}_p &=& \boldsymbol{v}_n + \boldsymbol{a}_n \Delta t
	+ \frac{\boldsymbol{j}_n}{2} \Delta t^2, 
\end{eqnarray}
where $\boldsymbol{x}_p$ and $\boldsymbol{v}_p$ are the predicted position
and velocity at the new time, $t_{n+1} = t_{n} + \Delta t$,
$\boldsymbol{x}_n$ and $\boldsymbol{v}_n$ are the
position and velocity at time $t_n$, and $\boldsymbol{a}_n$ and
$\boldsymbol{j}_n$ are the acceleration and jerk (first time derivative of
acceleration) at time $t_n$.
Using $\boldsymbol{x}_p$ and $\boldsymbol{v}_p$, we can now calculate the
acceleration and jerk, $\boldsymbol{a}_{n+1}$ and $\boldsymbol{j}_{n+1}$,
at time $t_{n+1}$.
Using $\boldsymbol{a}_n$, $\boldsymbol{j}_n$, $\boldsymbol{a}_{n+1}$ and
$\boldsymbol{j}_{n+1}$, we can construct the third-order Hermite
interpolation polynomial for $\boldsymbol{a}(t)$ as 
\begin{eqnarray}
	\label{eq:hermite,acc}
	\boldsymbol{a}(t) &=& \boldsymbol{a}_n + \boldsymbol{j}_n (t - t_n)
	+ \frac{\boldsymbol{s}_n}{2}(t-t_n)^2 + \frac{\boldsymbol{c}_n}{6}(t-t_n)^3,
\end{eqnarray}
where $\boldsymbol{s}_n$ and $\boldsymbol{c}_n$ are given by
\begin{eqnarray}
	\label{eq:hermite,snap}
	\boldsymbol{s}_n &=& \frac{-6(\boldsymbol{a}_n - \boldsymbol{a}_{n+1}) - \Delta t(4\boldsymbol{j}_n + 2\boldsymbol{j}_{n+1})}{\Delta t^2}, \\
	\label{eq:hermite,crackle}
	\boldsymbol{c}_n &=& \frac{12(\boldsymbol{a}_n - \boldsymbol{a}_{n+1}) + 6\Delta t(\boldsymbol{j}_n + \boldsymbol{j}_{n+1})}{\Delta t^3}.
\end{eqnarray}
We integrate equation (\ref{eq:hermite,acc}) from $t_{n}$ to
$t_{n+1}$ and obtain correctors given by
\begin{eqnarray}
	\label{eq:x,corrector}
	\boldsymbol{x}_c &=& \boldsymbol{x}_p + \frac{\boldsymbol{s}_n}{24} \Delta t^4 + \frac{\boldsymbol{c}_n}{120} \Delta t^5, \\
	\label{eq:v,corrector}
	\boldsymbol{v}_c &=& \boldsymbol{v}_p + \frac{\boldsymbol{s}_n}{6} \Delta t^3 + \frac{\boldsymbol{c}_n}{24} \Delta t^4.
\end{eqnarray}
If we set $\boldsymbol{x}_{n+1} = \boldsymbol{x}_c$ and
$\boldsymbol{v}_{n+1} = \boldsymbol{v}_c$ at this point, that means we use
the PEC (predict-evaluate-correct) form of the linear multistep method.
We can also use PECE or P(EC)$^2$ forms.

\subsection{Derivation of high-order time derivatives for hydrodynamical equations}
\label{sec:form_base}
In this section, we describe how we calculate high-order time derivatives
for hydrodynamics equations in Lagrangian view.
Our approach is essentially the same as that of \citet{Aoki1997IDO}, who
derived higher-order time derivatives for Eulerian view.
\citet{Aoki1997IDO} considered the following equation:
\begin{eqnarray}
	\label{eq:eq1,aoki1997}
	\frac{\partial}{\partial t}f = \xi_x f,
\end{eqnarray}
where $\xi_x$ is some linear operator. By taking time derivatives of
both sides of equation (\ref{eq:eq1,aoki1997}), they derived a series of
equations,
\begin{eqnarray}
	\frac{\partial^2}{\partial t^2} f &=& \xi_x\xi_x f, \\
	\frac{\partial^3}{\partial t^3} f &=& \xi_x\xi_x\xi_x f,
\end{eqnarray}
and so on. In this paper, we consider the equation
\begin{eqnarray}
	\label{eq:my,eq1,aoki1997}
	\frac{d}{d t}f = \xi_x f,
\end{eqnarray}
where $d/dt$ is the Lagrangian derivative,
\begin{eqnarray}
	 \frac{d}{dt} = \frac{\partial}{\partial t} + \boldsymbol{v}\cdot\boldsymbol{\nabla}.
\end{eqnarray}
The original set of partial differential equations of a Lagrangian
formulation of hydrodynamics is given by
\begin{eqnarray}
	 \label{eq:fluid,orig,eoc}
	 \frac{d\rho}{dt}
	 &=& -\rho \boldsymbol{\nabla}\cdot\boldsymbol{v}, \\
	 \label{eq:fluid,orig,eom}
	 \frac{d\boldsymbol{v}}{dt}
	 &=& -\frac{\boldsymbol{\nabla}P}{\rho}, \\
	 \label{eq:fluid,orig,eoe}
	 \frac{du}{dt}
	 &=& -\frac{P}{\rho} \boldsymbol{\nabla}\cdot\boldsymbol{v}, \\
	 \label{eq:fluid,orig,eos}
	 P &=& P(\rho, u).
\end{eqnarray}
Here, we rewrite $(d/dt)(\boldsymbol{\nabla})$ as
\begin{eqnarray}
	\label{eq:oper,t,x,org}
	\frac{d}{dt}\boldsymbol{\nabla} = \boldsymbol{\nabla}\frac{d}{dt} - \boldsymbol{\Diamond}.
\end{eqnarray}
The operator $\boldsymbol{\Diamond}$ is defined as
\begin{eqnarray}
	{\Diamond}_{\alpha} = (\nabla_\alpha v_\beta)(\nabla_\beta),
\end{eqnarray}
where $\alpha$ and $\beta$ are indices of dimensions, and 
\begin{eqnarray}
	\nabla_\alpha = \frac{\partial}{\partial x_\alpha},
\end{eqnarray}
where $\alpha =$ $1$, $2$ and $3$, and $\boldsymbol{x} = (x_1, x_2, x_3) = (x, y, z)$. The index $\beta$ is
summed over. Second time derivatives of $\rho$, $\boldsymbol{v}$ and
$u$ are then expressed as
\begin{eqnarray}
	 \frac{d^2\rho}{dt^2}
	 &=& \rho (\boldsymbol{\nabla}\cdot\boldsymbol{v})^2 + \rho \boldsymbol{\Diamond}\cdot\boldsymbol{v}
	 + \Delta P - \frac{\boldsymbol{\nabla}\rho\cdot\boldsymbol{\nabla}P}{\rho},\\ 
	 \frac{d^2\boldsymbol{v}}{dt^2}
	 &=& \frac{1}{\rho}\boldsymbol{\nabla}
	 \left[\widetilde{P}(\boldsymbol{\nabla}\cdot\boldsymbol{v})\right] + \frac{\boldsymbol{\Diamond}P}{\rho}
	 - \frac{(\boldsymbol{\nabla}\cdot\boldsymbol{v})(\boldsymbol{\nabla}P)}{\rho},\\
	 \frac{d^2u}{dt^2}
	 &=& \left(\frac{\widetilde{P} - P}{\rho}\right)(\boldsymbol{\nabla}\cdot\boldsymbol{v})^2
	 + \frac{P\Delta P}{\rho^2} - \frac{P(\boldsymbol{\nabla}P)\cdot(\boldsymbol{\nabla}\rho)}{\rho^3}
	 + \frac{P\boldsymbol{\Diamond}\cdot\boldsymbol{v}}{\rho},
\end{eqnarray}
where $\widetilde{P}$ is defined as
\begin{eqnarray}
	\label{eq:def,tildeP}
	\widetilde{P} \equiv \frac{P}{\rho}\frac{\partial P}{\partial u} + \rho\frac{\partial P}{\partial \rho}.
\end{eqnarray}
For the equation of state for ideal gas used in section
\ref{sec:test_shock},
\begin{eqnarray}
	\label{eq:eos,idealgas}
	P = (\gamma - 1) \rho u,
\end{eqnarray}
where $\gamma$ is the ratio of specific heat,
$\widetilde{P}$ is given by
\begin{eqnarray}
	\widetilde{P} = \gamma P.
\end{eqnarray}
For the equation of state for weakly compressible fluid used in section
\ref{sec:test_gravwave},
\begin{eqnarray}
	\label{eq:eos,weak}
	P = c_0^2(\rho - \rho_{\mathrm{air}}) + P_{\mathrm{air}},
\end{eqnarray}
where $\rho_{\mathrm{air}}$, $P_{\mathrm{air}}$, $g$, $H$ and $c_0$ are air density,
air pressure, gravity, height of fluid and sound velocity.
We set
\begin{eqnarray}
	\label{eq:cs,weak}
	c_0 = \sqrt{gH}.
\end{eqnarray}
The parameter $\widetilde{P}$ is given by
\begin{eqnarray}
	\widetilde{P} = c_0^2 \rho.
\end{eqnarray}
In this paper, we apply artificial viscosity of the form the same as that
in \citet{2017arXiv170105316Y}.
Note that we do not calculate the  contribution of the artificial
viscosity to the second time derivatives since artificial viscosity is not
differentiable. Therefore, the artificial viscosity for PEC and
P(EC)$^\infty$ forms of Hermite schemes are integrated with the Heun's
scheme and the trapezoidal scheme, respectively.
We calculate artificial viscosity as follows.
\begin{eqnarray}
	\label{eq:av,eom}
	\frac{d\boldsymbol{v}}{dt} &=& -\frac{\boldsymbol{\nabla}{q}}{\rho}, \\
	\label{eq:av,eoe}
	\frac{du}{dt} &=& -\frac{q}{\rho}\boldsymbol{\nabla}\cdot\boldsymbol{v}, \\
	\label{eq:av,q}
	q &=& -\left(\frac{|\sum_m\lambda_m|}{\sum_m|\lambda_m|}\right)^2\zeta
	\left[\alpha_{\mathrm{AV}}\rho c_s h_{\mathrm{AV}} + \beta_{\mathrm{AV}}\rho h_{\mathrm{AV}}^2 |\lambda_{\mathrm{mmax}}|\right]
	\lambda_{\mathrm{mmax}}\Theta(-\boldsymbol{\nabla}\cdot\boldsymbol{v}),
\end{eqnarray}
where $\alpha_{\mathrm{AV}}$, $\beta_{\mathrm{AV}}$ and $h_{\mathrm{AV}}$ are
coefficients, and $c_s$ and $\zeta $ are the sound velocity and
a parameter which controls the overall strength of AV.
In this paper, we set
$\alpha_{\mathrm{AV}} = 1$ and $\beta_{\mathrm{AV}} = 2$. The parameters $\lambda_{m}$ are
the eigenvalues of the strain rate tensor $\boldsymbol{\mathrm{s}}$ defined as
\begin{eqnarray}
	s_{\alpha, \beta} = \frac{1}{2}
	\left(\frac{\partial v_{\alpha}}{\partial x_{\beta}} + \frac{\partial v_{\beta}}{\partial x_{\alpha}} \right).
\end{eqnarray}
The parameter $\lambda_{\mathrm{mmax}}$ is the negative eigenvalue with the maximum
absolute value. If all eigenvalues are non-negative, $q=0$.
In this paper, we use the time-independent coefficient $\zeta$.
We set $\zeta = 1$.

\subsection{Calculation cost for time high order derivatives}
For the fourth-order Hermite time integrations, we must derive
second spatial order derivatives of physical quantities to
calculate jerk, snap and crackle.
However, if we use spatial high-order mesh-free methods (e.g.,
CPHSF), the additional number of arithmetic operations of jerk,
snap and crackle is much smaller than the original number of
the calculations of the spatial high-order mesh-free method.

In this section, we compare the original number of arithmetic
operations and the additional number of the operations necessary
for the Hermite scheme.
First, we show how to derive the spatial high-order derivatives
of a physical quantity $f$.
Second, the original number of arithmetic operations of CPHSF
is derived.
We call this value $N_{\rm op}$.
Note that we assume that $N_{\rm op}$
comprises only the
number of operations for the
evaluation of the inverse matrix of $B_i$ in equation
(\ref{eq:df,cphsf}) and interaction calculation between
particles since these dominate the total calculation cost of CPHSF.
Thirdly, the additional number of arithmetic operations for
jerk, snap and crackle is derived.
We call this value $N_{\rm add}$.
Finally, we compare $N_{\rm op}$ and $N_{\rm add}$.
To obtain the number of arithmetic operations, we calculate the
number of floating-point operations per one particle of CPHSF.
If a quantity have been derived, we assume that it will not be
unnecessarily recalculated. We assume that the numbers of
floating-point
operations required to evaluate division and square root are both 20.

First, we show how to derive the spatial high-order derivatives of
$f$.
In CPHSF, the $m$-th spatial order derivatives of $f$ is given
by the following equations,
\begin{eqnarray}
	\label{eq:df,cphsf}
	\delta^{m} f &=& \sum_{\alpha}
	\left[B^{-1}_i\right]_{m\alpha}
	\sum_j f_j p_{\alpha,ij}W_{ij}, \\
	\label{eq:delta,cphsf}
	\boldsymbol{\delta} &=& \left(1,\nabla_x, \nabla_y, \nabla_z, \frac{1}{2}\nabla_x^2, \nabla_x\nabla_y, \dots, \nabla_y\nabla_z^{n_p-1}, \nabla_z^{n_p}\right)^{T}, \\
	\label{eq:pij,cphsf}
	\boldsymbol{p}_{ij} &=& \left(1,x_{ij}, y_{ij}, z_{ij}, x_{ij}^2, x_{ij}y_{ij}, \dots, y_{ij}z_{ij}^{n_p-1}, z_{ij}^{n_p}\right)^{T}, \\
	\label{eq:Bi,cphsf}
	B_i &=& \sum_j W_{ij} \boldsymbol{p}_{ij} \otimes \boldsymbol{p}_{ij},
\end{eqnarray}
where $i$ and $j$ are indices of particles, $m$ and $\alpha$ are integers,
$n_p$ and $W_{ij}$ are the spatial order of the scheme and a Kernel
function and $x_{ij}$, $y_{ij}$ and $z_{ij}$ are $x_j-x_i$, $y_j-y_i$ and
$z_j-z_i$.

In CPHSF, the total number of floating point operations per one neighbor
particle is given by
\begin{eqnarray}
	N_{\rm op} = N_{\rm int}N_{\rm nb} + N_{\rm inv},
\end{eqnarray}
where $N_{\rm nb}$ is the number of neighbor particles, and $N_{\rm int}$
and $N_{\rm inv}$ are the numbers of floating-point operations for interaction
calculation between particles and the evaluation of the inverse matrix of
$B_i$ in equation (\ref{eq:df,cphsf}).
The number of floating-point operations for interaction calculation is 
given by
\begin{eqnarray}
	N_{\rm int} = N_{\rm dist} + N_{\rm kernel} + N_{\rm sf},
\end{eqnarray}
where $N_{\rm dist}$ and $N_{\rm kernel}$ are the number of floating-point
operations necessary to evaluate the relative distance and the kernel
function. The last term, $N_{\rm sf}$, represent the number of
floating-point operations for the CPHSF fitting. 
In CPHSF, first of all, we evaluate only $|\boldsymbol{x}_{ij}|/h_i$,
where $\boldsymbol{x}_{ij}$ is the displacement of particle $i$
and particle $j$ and $h_i$ is the Kernel length of particle $i$,
to search neighbor particles of particle $i$, and $N_{\rm dist}$ are
$\simeq 22$, $45$ and $48$ for 1, 2 and 3 dimensions.
Then, we evaluate elements of $B_i$ given by equation
(\ref{eq:Bi,cphsf}), polynomial equation given by equation
(\ref{eq:pij,cphsf}) and kernel function $W_{ij}$ to calculate
equation (\ref{eq:df,cphsf}).
One interaction calculation between particle $i$ and particle $j$ in
$[B_i]_{\alpha\beta}$ is given by
$\left\{[p_{ij}]_{\alpha}[p_{ij}]_{\beta}W_{ij}\right\}$.
The number of combinations of $[p_{ij}]_{\alpha}[p_{ij}]_{\beta}$
is $n(2n_p,D)$.
The parameter $n(n_p,D)$ is the number of bases of a polynomial fitting
in equation (\ref{eq:df,cphsf}), where $D$ is the number of dimensions,
and the value of $n(n_p,D)$ is given by
\begin{eqnarray}
	n(n_p,D) = \frac{1}{D!}\prod_{m = 0}^{D-1} (n_p+m).
\end{eqnarray}
For example, if we consider one dimensional case,
$\left\{[p_{ij}]_{\alpha}[p_{ij}]_{\beta}W_{ij}\right\}$ is given by
$x_{ij}^\alpha x_{ij}^\beta W_{ij}$ and thus
$[B_i]_{\alpha_1\beta_1}$ is the same as $[B_i]_{\alpha_2\beta_2}$ with
$(\alpha_1 + \beta_1) = (\alpha_2 + \beta_2)$.
Therefore, the number of the terms of the form of
$\left\{[p_{ij}]_{\alpha}[p_{ij}]_{\beta}W_{ij}\right\}$ is $n(2n_p,D)$.
Since we assume that a quantity, which have been derived, will not be
unnecessarily recalculated, the number of floating-point operations
for the evaluation of
$\left\{[p_{ij}]_{\alpha}[p_{ij}]_{\beta}W_{ij}\right\}$ except for
$\left\{[p_{ij}]_0[p_{ij}]_0W_{ij}\right\}$ is 1.
For example, if we consider one dimensional case,
we can get $x_{ij}^{m}W_{ij}$ by multiplying $x_{ij}^{m-1}W_{ij}$ by
$x_{ij}$ and thus the number of floating-point operations is only 1 for
the evaluation of $x_{ij}^{m}W_{ij}$.
In addition, the number of floating-point operations for summing
each term $\left\{[p_{ij}]_{\alpha}[p_{ij}]_{\beta}W_{ij}\right\}$
with respect to $j$ is 1.
Therefore, the total number of floating-point operations for one interaction
calculation in $B_i$ is $2n(2n_p,D)-1$.
One interaction calculation between particle $i$ and particle $j$ in
the calculation of equation (\ref{eq:df,cphsf}) is given by
${W}_{ij}f_j\boldsymbol{p}_{ij}$.
The number of the terms of the form of ${W}_{ij}f_j[p_{ij}]_{\alpha}$ is
$n(n_p,D)$.
We have density (pressure), energy and velocity, and thus
the number of physical quantities is $(D+2)$.
Therefore, the total number of floating-point operations for one
interaction calculation in $m$-th derivatives of density (pressure),
energy and velocity given equation (\ref{eq:df,cphsf}) is $2(D+2)n(n_p,D)$.
Therefore, the number of floating-point operations for the
CPHSF fitting is given by
\begin{eqnarray}
	N_{\rm sf}(n_p,D) = 2n(2n_p,D)+2(D+2)n(n_p,D) - 1
\end{eqnarray}
The numbers of floating-point operations that is necessary to evaluate
the kernel function, $N_{\rm kernel}$ are $\simeq 33$, $35$ and $36$ for
1, 2 and 3 dimensions.
From the above, the total numbers of floating-point operations for the
calculation of equation (\ref{eq:df,cphsf}) are $\simeq[33+N_{\rm sf}(n_p,1)]$,
$\simeq[35+N_{\rm sf}(n_p,2)]$ and $\simeq[36+N_{\rm sf}(n_p,3)]$ for 1, 2
and 3 dimensions.

From the above, the total numbers of floating-point operations for one
interaction calculation of CPHSF, $N_{\rm int}$, are
\begin{eqnarray}
N_{\rm int}\simeq[55+N_{\rm sf}(n_p,1)], \\
N_{\rm int}\simeq[80+N_{\rm sf}(n_p,2)], \\
N_{\rm int}\simeq[84+N_{\rm sf}(n_p,3)],
\end{eqnarray}
for 1, 2 and 3 dimensions.
Table \ref{tab:NNnbcalculationcost,cphsf} shows the summary of the numbers
of floating-point operations for one interaction calculation of CPHSF.

\begin{table}[htb]
	\begin{center}
	\caption{The numbers of floating-point operations for one interaction calculation of CPHSF}
	\scalebox{1.0}[1.0]{
	\begin{tabular}{c|ccc}
		Process           & $D=1$                 & $D=2$                & $D=3$                \\ \hline
		$N_{\rm dist}$    & $\simeq$22            & $\simeq$45           & $\simeq$48           \\
		$N_{\rm kernel}$  & $\simeq$33            & $\simeq$35           & $\simeq$36           \\ 
		$N_{\rm sf}$      & $N_{\rm sf}(n_p,1)$   & $N_{\rm sf}(n_p,2)$  & $N_{\rm sf}(n_p,3)$  \\ \hline
		Total             & $\simeq[55+N_{\rm sf}(n_p,1)]$ & $\simeq[80+N_{\rm sf}(n_p,2)]$  & $\simeq[84+N_{\rm sf}(n_p,3)]$  \\
	\end{tabular}
	}
	\label{tab:NNnbcalculationcost,cphsf}
	\end{center}
\end{table}
The evaluation of the inverse matrix of $B_i$ also dominates in
CPHSF and the number of floating-point operations of it, $N_{\rm inv}$,
is $\simeq 2n(n_p,D)^{3}/3$.
Therefore, the numbers of floating-point
operations per one particle of CPHSF are
\begin{eqnarray}
N_{\rm op} \simeq N_{\rm nb}[55+N_{\rm sf}(n_p,1)] + \frac{2}{3}n(n_p,1)^{3}, \\
N_{\rm op} \simeq N_{\rm nb}[80+N_{\rm sf}(n_p,2)] + \frac{2}{3}n(n_p,2)^{3}, \\
N_{\rm op} \simeq N_{\rm nb}[84+N_{\rm sf}(n_p,3)] + \frac{2}{3}n(n_p,3)^{3},
\end{eqnarray}
for 1, 2 and 3 dimension.

In the following, we derive $N_{\rm add}$.
To derive jerk, snap and crackle in the Hermite schemes,
we need to calculate second spatial order derivatives of $f_i$
given by equation (\ref{eq:df,cphsf}).
Here, the values of $\sum_j f_j p_{\alpha,ij}W_{ij}$ and
$\left[B^{-1}_i\right]_{m\alpha}$ have been calculated in the
derivation of the spatial first order derivative.
Therefore, we must calculate only the multiplication of
$\left[B^{-1}_i\right]_{m\alpha}$ by $\sum_j f_j
p_{\alpha,ij}W_{ij}$ and the additional number of calculations
for one physical quantity is given by $_{D}{\rm H}_2 n(n_p,D)$.
We have density (pressure), energy and velocity,
and thus the number of physical quantities in a numerical
calculation is $(D+2)$.
Therefore, the total additional number of calculations is
$N_{\rm add} = (D+2)_{d}{\rm H}_2 n(n_p,D)$.

\begin{figure}[ht]
	\begin{center}
	\includegraphics[width=12cm,clip]{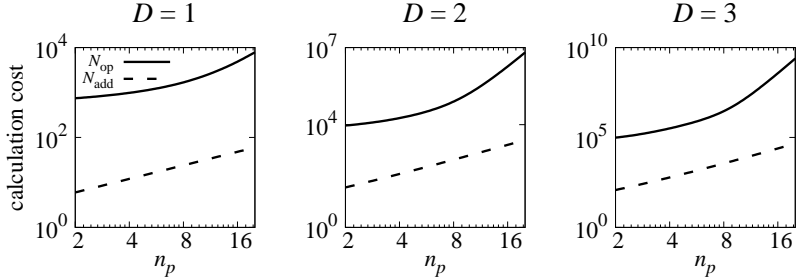} 
	\end{center}
	\caption{The panels show the values of $N_{\rm add}$
		and $N_{\rm op}$ plotted against $n_p$.
		The dashed and solid lines show these values for
		$N_{\rm add}$ and $N_{\rm op}$. From left to right,
		the values of $D$ are 1, 2 and 3.
	}
	\label{fig:calccost}
\end{figure}

Figure \ref{fig:calccost} shows $N_{\rm op}$ and $N_{\rm add}$ with
respect to $n_p$. 
We assume $N_{\rm nb} =$ 10, 75 and 600 for 1, 2 and 3 dimensions.
We can see that $N_{\rm add}$ is much smaller than $N_{\rm op}$.
Therefore, we conclude that the additional number of the calculations
of jerk, snap and crackle is much smaller than the original number of
the calculations of CPHSF.

\section{Numerical Experiments}
\label{sec:test}
In this section, we present the result of the Sod shock tube test in
section \ref{sec:test_shock} and that for the test of the surface gravity
wave in section \ref{sec:test_gravwave}.
We compare the results of fourth-order Hermite schemes and second and
fourth-order Runge-Kutta schemes in the Sod shock tube test, and the
results of an fully implicit Hermite-scheme, the implicit fourth-order
Runge-Kutta scheme and the backward-Euler scheme in the surface gravity
wave test.

\subsection{Sod shock tube}
\label{sec:test_shock}
In this section, we present the result of the Sod shock tube test
(\cite{sod1978survey}). We assume that fluid is ideal gas with
$\gamma = 1.4$. The computational domain is $-0.5 \leq x < 0.5$ with a
periodic boundary, and initial boundary of two fluids are at $x = -0.5$
and 0. In this test, we used equal-mass particles. Initial velocity is
given by $v_x = 0$. The density is smoothed by a $C^5$ polynomial, and is
given by
 \begin{eqnarray}
	\rho(x) = \left\{
	\begin{array}{ll}
		\rho_h  & -0.25 \leq x < -x_0, \\
		\frac{\rho_h - \rho_l}{2}\sum_{m=0}^5 b_m x^{2m+1} + \frac{\rho_h + \rho_l}{2}
		& -x_0 \leq x < x_0,\\
		\rho_l & x_0 \leq x \leq 0.25,
	\end{array}
	\right.
 \end{eqnarray}
where $(b_0, b_1, b_2, b_3, b_4, b_5) = (-693/256, 1155/256, -693/128,
495/128, -385/256, 63/256)$ and $\rho_h$ and $\rho_l$ are the values of
initial density in the high- and low-density regions. We used $\rho_h = 1$
and $\rho_l = 0.25$. The parameter $x_0$ represents the width of the
smoothing region, and we used two values of $x_0$. One is an initial
condition with $x_0 = 0.006$, and the other is a smooth initial condition
with $x_0 = 0.03$.
We set the initial condition for $0.25\leq x<0.5$ to be mirroring that of
$0<x\leq 0.25$,
and $-0.5\leq x \leq -0.25$ as mirroring $-0.25\leq x \leq 0$.
The positions of particles in the smoothing region are determined so
that position $x_i$ of particle $i$ satisfies
\begin{eqnarray}
	\label{eq:pos,shock}
	\int^{x_i}_{x_{i-1}} \rho(x) dx = \frac{1}{2N_h},
\end{eqnarray}
where $N_h$ is the number of particles in the high-density region and the
right-hand side of equation (\ref{eq:pos,shock}) is mass of a particle.
The smoothed pressure is given by
 \begin{eqnarray}
	P(x) = \left\{
	\begin{array}{ll}
		P_h & -0.25 \leq x < -x_0, \\
		\frac{P_h - P_l}{2}\sum_{m=0}^5 b_m x^{2m+1} + \frac{P_h + P_l}{2}
		& -x_0 \leq x < x_0,\\
		P_l & x_0 \leq x \leq 0.25,
	\end{array}
	\right.
\end{eqnarray}
where $P_h$ and $P_l$ are the values of initial pressure in the high- and
low-density regions. We used $P_h = 1$ and $P_l = 0.1795$. We used
equations (\ref{eq:av,eom}) and (\ref{eq:av,eoe}) for the artificial
viscosity with $h_{\mathrm{AV}} = 2.375\times10^{-3}$.
We used a sixth-order interpolation with the value of interpolation
polynomial at the position of particle $\boldsymbol{x}_i$ fixed to the
actual value.
Therefore, $\boldsymbol{\delta}$ given by equation (\ref{eq:delta,cphsf})
and $\boldsymbol{p}_{ij}$ given by equation (\ref{eq:pij,cphsf}) are
\begin{eqnarray}
	\boldsymbol{\delta} &=& \left(1, \nabla_x, \frac{1}{2!}\nabla_x^2, \frac{1}{3!}\nabla_x^3, \frac{1}{4!}\nabla_x^4, \frac{1}{5!}\nabla_x^5 \right)^{T}, \\
	\boldsymbol{p}_{ij} &=& \left(1,x_{ij}, x_{ij}^2, x_{ij}^3, x_{ij}^4, x_{ij}^5\right)^{T}.
\end{eqnarray}
The kernel function is the fourth-order Wendland function
(\cite{wendland1995piecewise}).
The kernel length is given by
\begin{eqnarray}
	\label{eq:kernellength}
	h_i &=& \eta \left(\frac{\tilde{m}_i}{\rho_i}\right)^{1/D}, \\
	\tilde{m}_i &=& \rho_{t=0,i} \Delta V_{t=0,i},
\end{eqnarray}
where $\rho_{t=0,i}$ and $\Delta V_{t=0,i}$ are density and
geometric volume of a particle $i$ at $t = 0$. We set $\eta = 3.8$.

We calculated L1-norm error of density at $t = 0.1$ to verify the
spatial order of the schemes and to compare the accuracy of the
schemes,
\begin{eqnarray}
	\label{eq:error,dens}
	\epsilon_{\rho} = \sum_{n=1}^{N_x}\frac{1}{N_x}
	\frac{|\rho_n - \rho_{n}^{\mathrm{hres}}|}{\rho_{n}^{\mathrm{hres}}},
\end{eqnarray}
where $\rho_{n}^{\mathrm{hres}}$ is the result of a high-resolution
test in which the number of particles, $N_x$, is 8000 and
$dt = 10^{-6}$. When we derived equation (\ref{eq:error,dens}), we
calculated $\rho_n$ of particles rearranged at positions same as those of
the high-resolution test. The time integrator for high-resolution test is
the Hermite scheme of the P(EC)$^2$ form.
For the test of the time order of the scheme for the test with $N_x = N_0$,
$\rho_{n,\Delta t}^{\mathrm{hres}}$ is the result of a high-resolution test in
which $N_x$ is $N_0$ and $dt = 10^{-6}$. The time integrator for
high-resolution test is the same as that for $\rho_n$. In this case we
define the error as
\begin{eqnarray}
	\epsilon_{\rho,\Delta t} = \sum_{n=1}^{N_x}\frac{1}{N_x}
	\frac{|\rho_n - \rho_{n,\Delta t}^{\mathrm{hres}}|}{\rho_{n,\Delta t}^{\mathrm{hres}}},
\end{eqnarray}

We compare results with PEC, PECE and P(EC)$^2$ forms of Hermite schemes,
and Heun's scheme (hereafter RK2) and the classical fourth-order Runge-Kutta
scheme (hereafter RK4). The numbers of particles, $N_x$, are $1000$, $2000$
and $4000$.

\begin{figure}[ht]
	\begin{center}
	\includegraphics[width=12cm,clip]{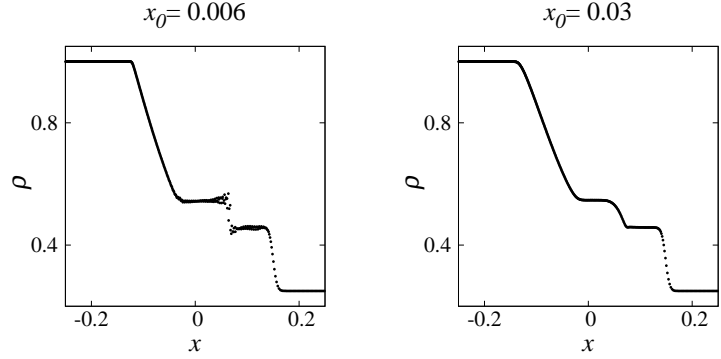} 
	\caption{
		The results of the Sod shock tube tests with $N_x = 1000$.
		The density profiles at $t = 0.1$ are shown.
		The left and right panels show the results for
		$x_0 = 0.006$ and $x_0 = 0.03$.
		}
		\label{fig:sodshockpro01N}
	\end{center}
 \end{figure}

\begin{figure}[ht]
	\begin{center}
	\includegraphics[width=12cm,clip]{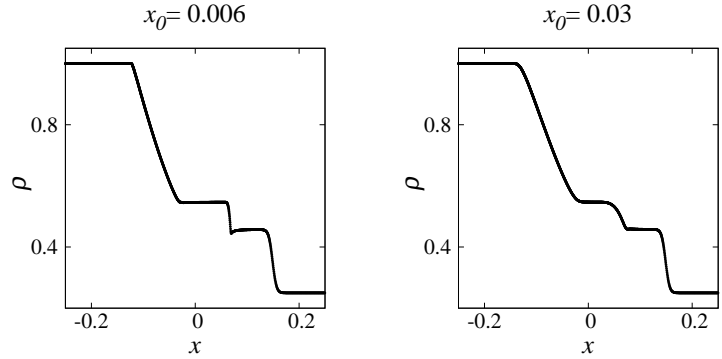} 
	\caption{
		The same as figure \ref{fig:sodshockpro01N}, but for
		$N_x=4000$.
		}
		\label{fig:sodshockpro04N}
	\end{center}
 \end{figure}

Figure \ref{fig:sodshockpro01N} shows density profiles at $t = 0.1$ for the
tests with $N_x = 1000$ and $dt \simeq dt_{\mathrm{max}}/4$, where
$dt_{\mathrm{max}}$ is the maximum time step in the stability region, with
the PEC form of the Hermite scheme.
Note that the results for all schemes are similar to that for the PEC form
of the Hermite scheme.
We can see that the shock wave can be captured. However, the post shock
oscillation is strong for $x_0 = 0.006$.
Figures \ref{fig:sodshockpro04N} is the same as figure
\ref{fig:sodshockpro01N}, but for $N_x = 4000$. Note that the results
are independent of the time integration scheme used and the results for
$N_x = 2000$ are similar to those for $N_x = 4000$.
We can see that the shock wave can be captured clearly even if initial
condition is not smooth.
Therefore, if the initial condition is not smooth, the resolution of time
and space should be higher.

\begin{figure}[ht]
	\begin{center}
		\includegraphics[width=6cm,clip]{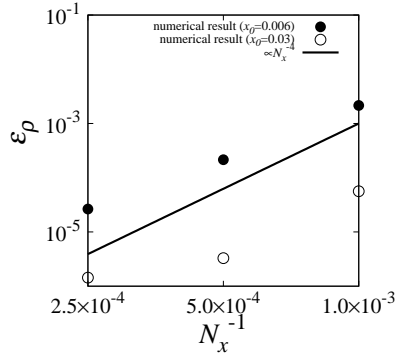} 
		\caption{
			The panel shows $\epsilon_{\rho}$ at
			$t=0.1$ for the tests with $x_0 = 0.006$ and $x_0 = 0.03$
			plotted against $N_x^{-1}$.
			Filled and open circles show results for $x_0=0.006$ and
			$0.03$, and solid curve shows the theoretical
			models for the error.
			}
		\label{fig:hermsodshockerrNN}
	\end{center}
\end{figure}

Now we check the spatial order of the scheme.
We used the sixth order shape function and then the first and second
derivatives are fifth and fourth orders in space.
Therefore, if the result converges to an exact solution following the order
of the method, the order of the scheme should be larger than or equal to four
and thus $\epsilon_{\rho}$ should be given by $\epsilon_{\rho} \propto
N_x^{-m}$ where $m$ is larger than or equal to 4.
Figure \ref{fig:hermsodshockerrNN} shows that $\epsilon_{\rho}$ for the
P(EC)$^2$ form of the Hermite scheme for runs with $dt = 10^{-6}$ plotted
against $N_x^{-1}$.
The results are independent of the time integration scheme used.
The value of $\epsilon_{\rho}$ for runs with $x_0 = 0.006$ is proportional
to $N_x^{-4}$.
The value of $\epsilon_{\rho}$ in the large $N_x$ region for runs with
$x_0 = 0.03$ is proportional to $N_x^{-1}$ since, in this region, the
round-off error dominates the total error.
In the other region, $\epsilon_{\rho}$ is proportional to $N_x^{-4}$.
From these results, we can conclude that the spatial order of the scheme
is consistent to theoretical expectation.

\begin{figure}[ht]
	\begin{center}
		\includegraphics[width=16cm,clip]{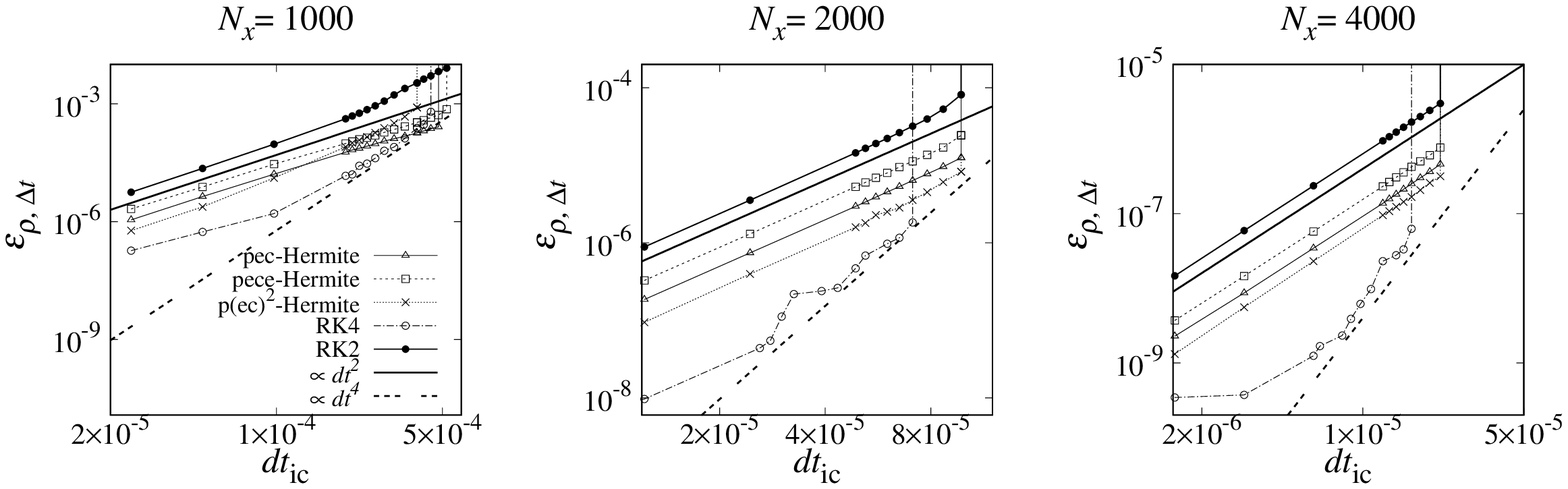} 
		\caption{
			The panel shows $\epsilon_{\rho, \Delta t}$ at $t=0.1$ for the tests
			with $x_0 = 0.006$ plotted against $dt_{\rm ic}$.
			From left to right panels, the results for the $N_x = 1000$,
			$2000$ and $4000$.
			The lower left and lower middle side panels show the
			results for the second and fourth Runge-Kutta schemes.
			Triangles, squares and crosses show the results for
			Hermite schemes in PEC, PECE, P(EC)$^2$ forms, and open
			and filled circles show the results for RK4 and RK2.
			Solid and dashed curves show
			the theoretical models for the error of second- and
			fourth-order schemes.
			}
		\label{fig:hermsodshock01errdt}
	\end{center}
\end{figure}

\begin{figure}[ht]
	\begin{center}
		\includegraphics[width=16cm,clip]{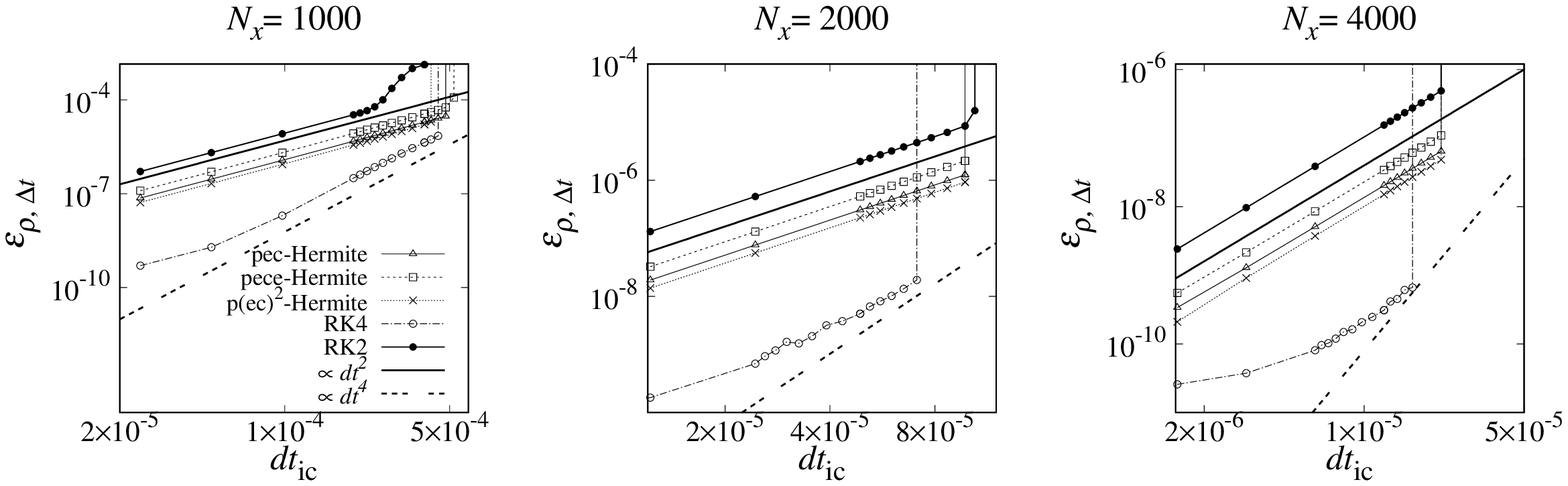} 
		\caption{
			Same as figure \ref{fig:hermsodshock01errdt}, but the
			results for $x_0 = 0.03$.
			}
		\label{fig:hermsodshock05errdt}
	\end{center}
\end{figure}

Let us look at the time orders of the schemes.
Figures \ref{fig:hermsodshock01errdt} and \ref{fig:hermsodshock05errdt}
show $\epsilon_{\rho,\Delta t}$ for the tests with $x_0 = 0.006$ and
$x_0 = 0.03$ plotted against $dt_{\mathrm{ic}}$ where $dt_{\mathrm{ic}}$
is $dt$ divided by the number of interaction calculations per one time
step. 
We can see that the errors of RK2 and RK4 are $\mathcal{O}(dt^2)$ and
$\mathcal{O}(dt^4)$ respectively, and that of the Hermite schemes are
$\mathcal{O}(dt^2)$.

In the following we explain the reason why the order of the Hermite scheme
is $\mathcal{O}(dt^2)$ for fixed $N_x^{-1}$. In a particle-based method,
the calculated spatial
derivatives contain discretization errors, and therefore the time
derivative contain errors. In the case of RK schemes, this error causes the
solution in the limit of $dt \rightarrow 0$ to converge to the solution
different from the exact solution, but the rate of the convergence is the
order of the time integration scheme, since we can regard the
space-discretized differential equations as the set of ordinal
differential equations.
However, in the case of the Hermite scheme, we construct the second time
derivatives of physical quantities from the original equations and
high-order spatial derivatives, and these spatial derivatives contain
discretization errors. Thus, both the first and second time derivatives
contain the errors due to space discretization errors, and therefore the
second time derivatives are not exactly the time derivatives of the first
time derivatives. For simplicity, let us illustrate this behaviour for the
integration of velocity in one dimension.
Here, we rewrite the correctors by substituting equations (\ref{eq:hermite,snap})
and (\ref{eq:hermite,crackle}) to equation (\ref{eq:v,corrector}).
Note that we set $dt = \Delta t$ in equation (\ref{eq:v,predictor}).
\begin{eqnarray}
	\label{eq:rew,v,predictor}
	{v}_c &=& {v}_n
	+ \frac{1}{2}({a}_{n} + {a}_{n+1})\Delta t
	+ \frac{1}{12}({j}_{n} - {j}_{n+1})\Delta t^2.
\end{eqnarray}
If we use sixth order polynomial fitting for deriving spatial
derivatives, ${v}_c$ containing the spatial errors is given by
\begin{eqnarray}
	\label{eq:xerr,v,predictor}
	{v}_c = {v}_n
	+ \frac{1}{2}\left({A}_{n} + {A}_{n+1}\right)\Delta t
	+ \frac{1}{12}\left({J}_{n} - {J}_{n+1}\right)\Delta t^2,
\end{eqnarray}
where $A_{n}$, $A_{n+1}$, $J_n$ and $J_{n+1}$ are accelerations at $n$ and
$n+1$ steps and jerks at $n$ and $n+1$ steps and are given by sixth order
polynomial fitting for deriving spatial derivatives.
Therefore, $J$ is not equal to the time derivative of $A$.
\begin{eqnarray}
	J = \frac{dA}{dt} + \epsilon_{J},
\end{eqnarray}
where $\epsilon_{J}$ is the error.
Here, we integrate equation (\ref{eq:xerr,v,predictor}) from $t=0$ to
$t=T$,
\begin{eqnarray}
	\label{eq:sum,v,predictor}
	{v}_{c,(t=T)} &=& {v}_{(t=0)}
	+ \sum_{n=0}^{N_t}\left[\frac{1}{2}\left({A}_{n} + {A}_{n+1}\right)\Delta t\right]
	+ \sum_{n=0}^{N_t}\left[\frac{1}{12}\left({J}_{n} - {J}_{n+1}\right)\Delta t^2\right] \nonumber \\
	&=& {v}_{(t=0)}
	+ \sum_{n=0}^{N_t}\left[\frac{1}{2}\left({A}_{n} + {A}_{n+1}\right)\Delta t\right]
	+ \sum_{n=0}^{N_t}\left[\frac{1}{12}\left(\frac{dA_n}{dt} + \epsilon_{J,n} - \frac{dA_{n+1}}{dt} - \epsilon_{J,n+1}\right)\Delta t^2\right],\nonumber\\
\end{eqnarray}
where $N_t$ is given by $N_t = T/\Delta t$.
Here, we can assume that 
\begin{eqnarray}
	\label{eq:vdt,A}
	\frac{dv}{dt} = \lim_{\Delta t = 0} A.
\end{eqnarray}
Therefore, equation (\ref{eq:sum,v,predictor}) becomes
\begin{eqnarray}
	{v}_{(t=T)} &=& {v}_{A}(T) + \mathcal{O}(\Delta t)^4
	+ \frac{1}{12}\left[\epsilon_{J,(t=0)} - \epsilon_{J,(t=T)}\right] \Delta t^2,
\end{eqnarray}
where $v_{A}(T)$ is the analytical solution for the velocity which
satisfies equation (\ref{eq:vdt,A}).
We can see that the time order of a Hermite scheme is equal to two.
From these results, we can conclude that the time orders of the schemes
are consistent.
The fact that the apparent error order of the Hermite scheme is two does
not imply it is a second order scheme, since when we simultaneously
shrink the interparticle distance and timestep, the error will be
$\mathcal{O}(dt^4)$ as expected. The second-order behaviour occurs only
when the spatial error dominates the total error.

\begin{figure}[ht]
	\begin{center}
	\includegraphics[width=16cm,clip]{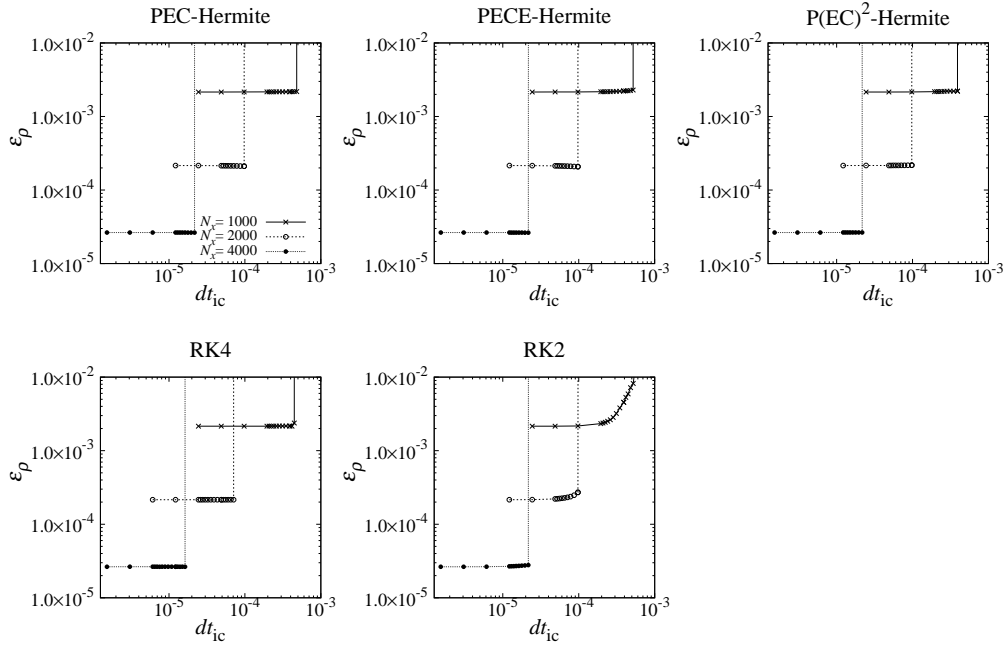} 
	\caption{
			The panel shows $\epsilon_{\rho}$ at $t=0.1$ for the tests
			with $x_0 = 0.006$ plotted against $dt_{\rm ic}$.
			From left to right in the upper panels, the results for the
			PEC-, PECE- and P(EC)$^2$ forms of the Hermite schemes are
			shown.
			The lower left and lower middle side panels show the
			results for the second and fourth Runge-Kutta schemes.
			Crosses, open and filled circles show results
			for $N_x=1000$, $2000$ and $4000$.
		}
		\label{fig:sodshockerr01s}
	\end{center}
 \end{figure}

Figure \ref{fig:sodshockerr01s} shows errors for tests with $x_0 = 0.006$
plotted against $dt_{\mathrm{ic}}$. The result shows that the accuracy of
fourth-order Hermite schemes is similar to those of RK2 and RK4, since the
errors of spatial differentiation approximation determines the overall
error. 

\begin{figure}[ht]
	\begin{center}
	\includegraphics[width=7cm,clip]{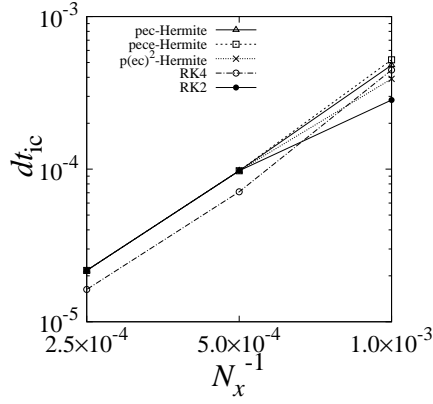} 
	\caption{
			The panels show maximum $dt_{\mathrm{ic}}$ in the numerical
			stable region for tests with $x_0 = 0.006$ plotted
			against $N_x^{-1}$.
			Triangles, squares and crosses show the results for
			Hermite schemes in PEC, PECE, P(EC)$^2$ forms, and open
			and filled circles show the results for RK4 and RK2.
		}
		\label{fig:sodshockLimitDt01s}
	\end{center}
 \end{figure}

Figure \ref{fig:sodshockLimitDt01s} shows maximum $dt_{\mathrm{ic}}$ in the
numerical stable region for tests with $x_0 = 0.006$ plotted against
$N_x^{-1}$.
We can see that the regions of stability of fourth-order Hermite
schemes are larger than or equal to those of RK2 and RK4.
Hence, we can use larger timesteps with the Hermite schemes.
Therefore, we can conclude that Hermite schemes, especially in PEC and
PECE forms, are better than Runge-Kutta schemes for simulations of
fluid with shock and contact discontinuity, even when the initial
condition has sharp jump.

\begin{figure}[ht]
	\begin{center}
	\includegraphics[width=16cm,clip]{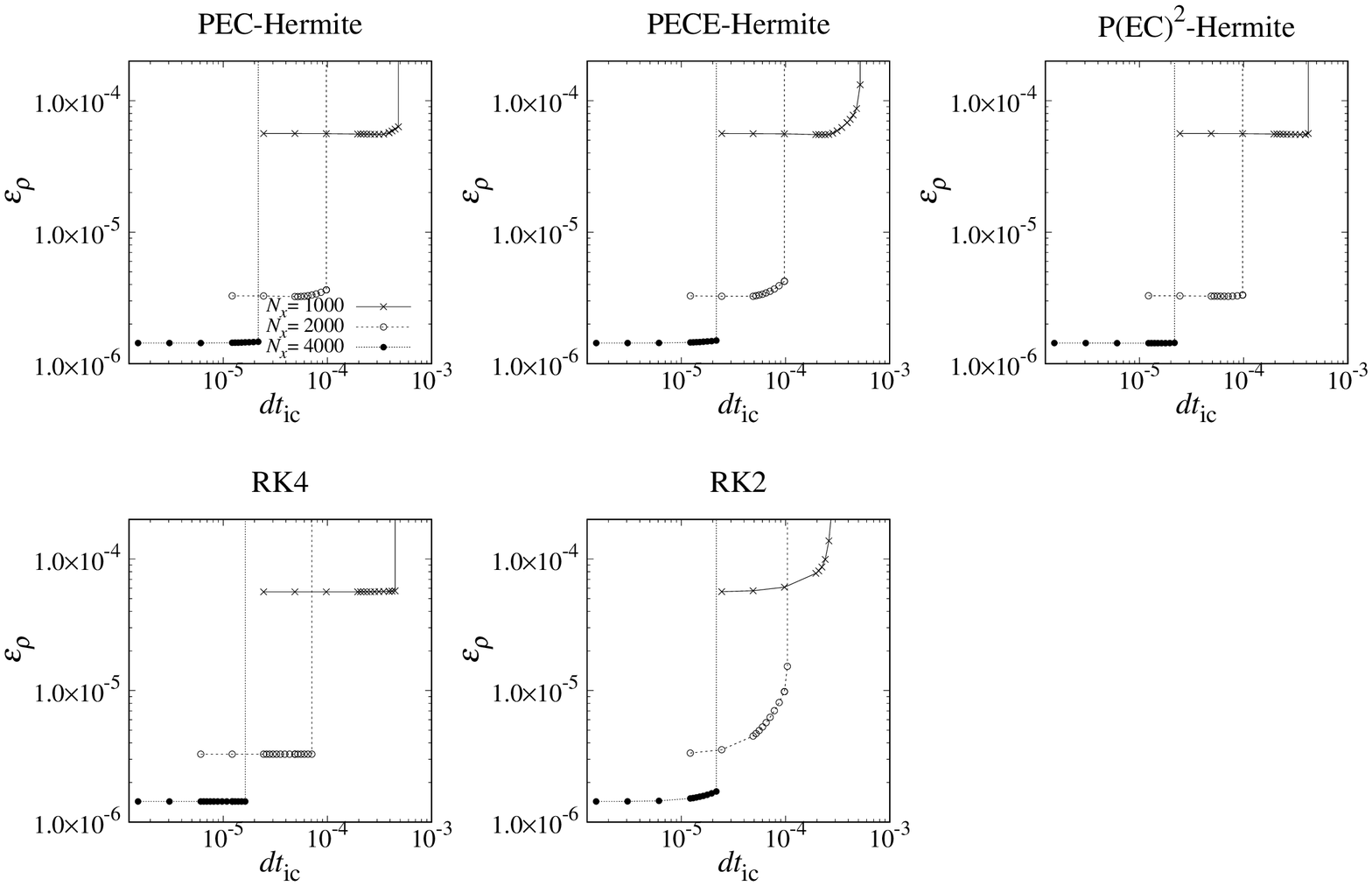} 
	\caption{
		The same as figure \ref{fig:sodshockerr01s}, but for
		$x_0 = 0.03$.
		}
		\label{fig:sodshockerr05s}
	\end{center}
 \end{figure}

Figure \ref{fig:sodshockerr05s} shows errors for $x_0 = 0.03$ plotted
against $dt_{\mathrm{ic}}$. As in the case of $x_0 = 0.006$, the results
show that the accuracy of fourth-order Hermite schemes is similar to
those of RK2 and RK4, since the errors of spatial differentiation
approximation determines the overall error. 

\begin{figure}[ht]
	\begin{center}
	\includegraphics[width=7cm,clip]{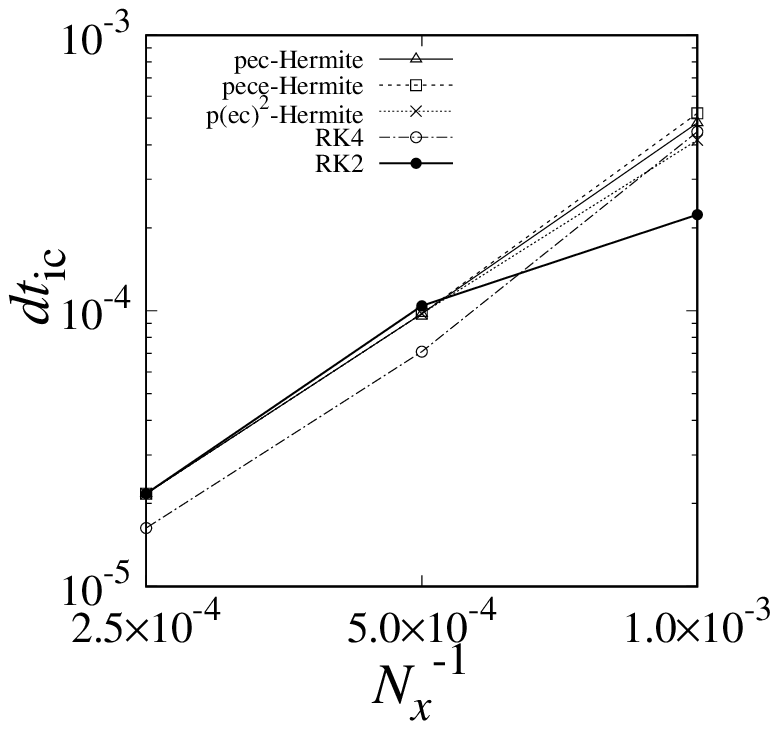} 
	\caption{
		The same as figure \ref{fig:sodshockLimitDt01s}, but for
		$x_0 = 0.03$.
		}
	\label{fig:sodshockLimitDt05s}
	\end{center}
 \end{figure}

Figure \ref{fig:sodshockLimitDt05s} shows maximum $dt_{\mathrm{ic}}$ in the
numerical stable region for tests with $x_0 = 0.03$ plotted against
$N_x^{-1}$.
As in the case of $x_0 = 0.006$, the results the regions of stability of
fourth-order Hermite schemes are larger than or equal to those
of RK2 and RK4.
Therefore, we can conclude that Hermite schemes, especially in PEC
and PECE forms, are better than Runge-Kutta schemes for simulations of
fluid with shock and contact discontinuity.
We can conclude that Hermite schemes are more computationally
efficient better than Runge-Kutta schemes for calculation shocks.




\subsection{Surface gravity wave test}
\label{sec:test_gravwave}
The surface gravity wave test is useful for the investigation of the
capability of numerical schemes to handle two-dimensional fluid dynamics
with high accuracy and small dissipation. The initial condition is the same
as those in \citet{2011CoPhC.182..866A} and \citet{2017arXiv170105316Y},
but sound velocity given by equation (\ref{eq:cs,weak}) is $10$
times smaller than that of \citet{2017arXiv170105316Y}. We assume that
fluid is weakly compressible with the equation of state given by equation
(\ref{eq:eos,weak}) with $\rho_{\mathrm{air}} = 10^3$ and $P_{\mathrm{air}} = 10^5$
and sound velocity given by equation (\ref{eq:cs,weak}) with $g = -10$ and
the height of fluid $H=1$. The computational domain is $0 \leq x < 1$,
$0 \leq y \leq 1$. We applied a periodic boundary at $x = 0$, $v_y = 0$ at
$y = 0$ and $P = P_{\mathrm{air}}$ for particles initially at $y = 1$ as
boundary conditions. Initial density is 
\begin{eqnarray}
	\rho(y) = \rho_{\mathrm{air}}e^{g(H-y)/c_0^2}.
\end{eqnarray}
Initial velocity is
\begin{eqnarray}
	v_x &=&  A\frac{|g|k}{\omega}\frac{\cosh(ky)}{\cosh(kH)}\sin(kx), \\
	v_y &=& -A\frac{|g|k}{\omega}\frac{\sinh(ky)}{\cosh(kH)}\cos(kx),
\end{eqnarray}
where $A$, $k$ and $\omega$ are the amplitude, the number of wave and its
frequency. We set $A = 0.01$, $k=2\pi$ and $\omega=\sqrt{|g|k\tanh(kH)}$.
In this test, we do not use artificial viscosity to clarify the origin of the
error. We used a fifth-order interpolation with the value of interpolate
polynomial at the position of particle $\boldsymbol{x}_i$ fixed to the
actual value.
Therefore, $\boldsymbol{\delta}$ given by equation (\ref{eq:delta,cphsf})
and $\boldsymbol{p}_{ij}$ given by equation (\ref{eq:pij,cphsf}) are
\begin{eqnarray}
	\boldsymbol{\delta} &=& \left(1, \nabla_x, \frac{1}{2!}\nabla_x^2, \nabla_x\nabla_y, \frac{1}{2!}\nabla_y^2, \dots, \frac{1}{2!3!}\nabla_x^2\nabla_y^3, \frac{1}{4!}\nabla_x\nabla_y^4, \frac{1}{5!}\nabla_y^5 \right)^{T}, \\
	\boldsymbol{p}_{ij} &=& \left(1,x_{ij}, y_{ij}, x_{ij}^2, x_{ij}y_{ij}, y_{ij}^2, \dots, x_{ij}^2y_{ij}^3, x_{ij}y_{ij}^4, y_{ij}^5\right)^{T}.
\end{eqnarray}
The kernel function is the fourth-order Wendland function
(\cite{wendland1995piecewise}). We used equation
(\ref{eq:kernellength}) as the kernel length and set $\eta = 3.8$.

We calculate the absolute error of $v_x$ at $(x,y) = (0.4,1)$ and
$t = 0.2T$ where $T$ is the period given by $2\pi/\omega$
for checking the spatial order of the schemes and comparing the
accuracy of the schemes.
\begin{eqnarray}
	\label{eq:error,vx}
	\epsilon_{v_x} = {|v_{x} - v_{x}^{\mathrm{hres}}|},
\end{eqnarray}
where $v_{x}^{\mathrm{hres}}$ is the result of the high-resolution test in
which the number of particles, $N$, is $128\times129$ and $dt = T/1024$.
The time integrator for high-resolution test is the implicit
Hermite scheme.
For checking the time order of the scheme for the test with $N = N_0$,
$v_{x,\Delta t}^{\mathrm{hres}}$ is the result of a high-resolution test in
which $N$ is $N_0$ and $dt = T/512$. The time integrator for
high-resolution test is same as $v_{x}$.
We calculated $v_x$ and $v_{x,\Delta t}^{\mathrm{hres}}$ of the particles
initially at $(x,y) = (0.3125,1)$.
In this case we define the error as
\begin{eqnarray}
	\epsilon_{v_x, \Delta t} = {|v_{x} - v_{x,\Delta t}^{\mathrm{hres}}|},
\end{eqnarray}

We compare results of runs with the implicit Hermite scheme, the
backward-Euler scheme (hereafter IRK1) and the Gauss-Legendre scheme
(hereafter IRK4). The numbers of particles, $N$, are $16\times17$,
$32\times33$ and $64\times65$.

\begin{figure}[ht]
	\begin{center}
	\includegraphics[width=8cm,clip]{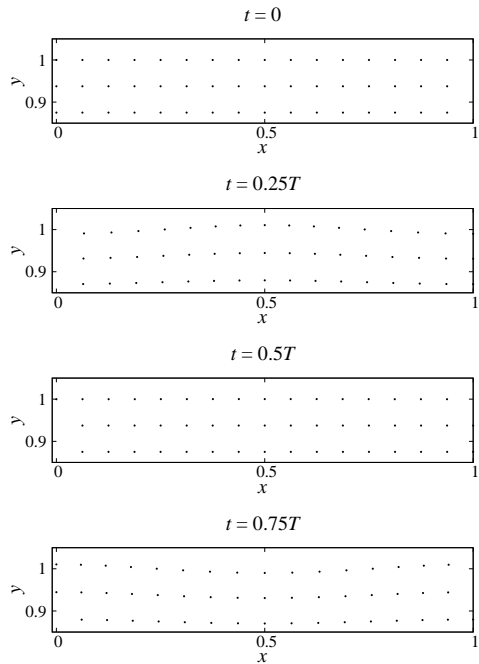} 
	\caption{
		Results of the surface gravity wave tests with
		$N = 16\times17$, form top to bottom, the snapshots at
		$t = 0$, $0.25T$, $0.5T$ and $0.75T$ are shown.
		}
		\label{fig:gravwavepro16Nx}
	\end{center}
 \end{figure}
\begin{figure}[ht]
	\begin{center}
	\includegraphics[width=12cm,clip]{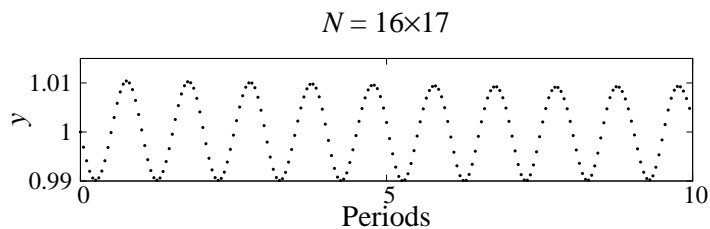} 
	\caption{
		Time evolution of the $y$-coordinate of the particle
		initially at $(x,y) = (0,1)$ in the surface gravity wave
		test with $N = 16\times17$.
		}
		\label{fig:gravwavepos16Nx}
	\end{center}
 \end{figure}

Figure \ref{fig:gravwavepro16Nx} shows the time evolution up to
$t = 0.75T$ with the implicit Hermite scheme, $N = 16 \times 17$ and
$dt \simeq dt_{\mathrm{max}}/4$. Figure \ref{fig:gravwavepos16Nx} shows
$y$ of the particle initially at $(x,y) = (0,1)$ with the implicit Hermite
scheme, $N = 16 \times 17$ and $dt \simeq dt_{\mathrm{max}}/4$.
Note that the results are independent of the time integration scheme used
and $N$.

\begin{figure}[ht]
	\begin{center}
		\includegraphics[width=6cm,clip]{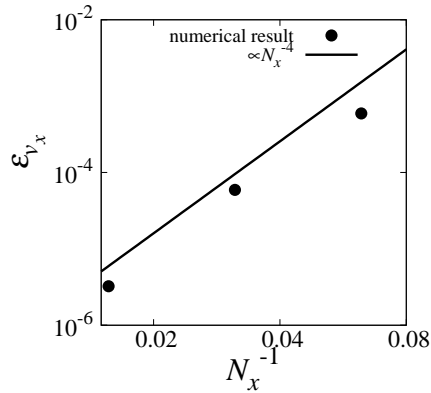} 
		\caption{
			The panel shows $\epsilon_{v_x}$ at $t=0.2T$ plotted
			against $N_x^{-1}$.
			Filled circles show numerical results and solid curves
			show the theoretical models for the error.
			}
		\label{fig:gravwaveerrNN}
	\end{center}
\end{figure}

Now we check the spatial order of the scheme.
We used the fifth order shape function and then the first and second
derivatives are fourth and third orders in space.
Therefore, if the result converges to an exact solution following the order
of the method, the order of the scheme should be larger than or equal to
three and thus $\epsilon_{v_x}$ should be given by $\epsilon_{v_x} \propto
N_x^{-m}$ where $m$ is larger than or equal to 3 and $N_x$ is
the number of particles of the $x$-direction.
Figure \ref{fig:gravwaveerrNN} shows $\epsilon_{v_x}$ for the implicit
Hermite scheme with $dt = T/512$ plotted against $N_x^{-1}$.
We can see that the error $\epsilon_{v_x}$ is proportional to $N_x^{-4}$.
Therefore, the error in acceleration determines the overall error.
The results are independent of the time integration used.
From the result, the spatial order of the scheme is consistent.

\begin{figure}[ht]
	\begin{center}
		\includegraphics[width=16cm,clip]{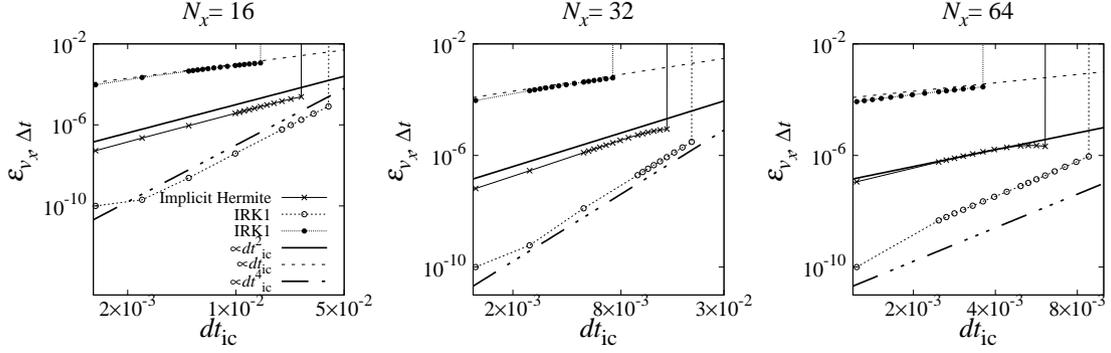} 
		\caption{
			The panel shows $\epsilon_{v_x, \Delta t}$ plotted against
			$dt_{\rm ic}$.
			From left to right panels, the results for the $N_x = 16$,
			$32$ and $64$.	
			Crosses, open and filled circles show the results of the
			implicit Hermite scheme, IRK4 and IRK1.
			Dashed, solid and dot-dot-dot dashed curves show
			the theoretical models for the error of second-, first-
			and fourth-order schemes.
			}
		\label{fig:hermgravewaveerrdt}
	\end{center}
\end{figure}

Let us now look at the time order of the scheme.
Figure \ref{fig:hermgravewaveerrdt} shows that $\epsilon_{v_x, \Delta t}$ plotted
against $dt_{\mathrm{ic}}$.
We can see that the errors of the implicit Hermite scheme, IRK4 and IRK1
are $\mathcal{O}(dt^2)$, $\mathcal{O}(dt^4)$ and $\mathcal{O}(dt)$
respectively.
As described in section \ref{sec:test_shock}, the time order of Hermite
scheme is equal to two.
From these results, we can conclude that the time orders of the schemes
are consistent.

\begin{figure}[ht]
	\begin{center}
	\includegraphics[width=16cm,clip]{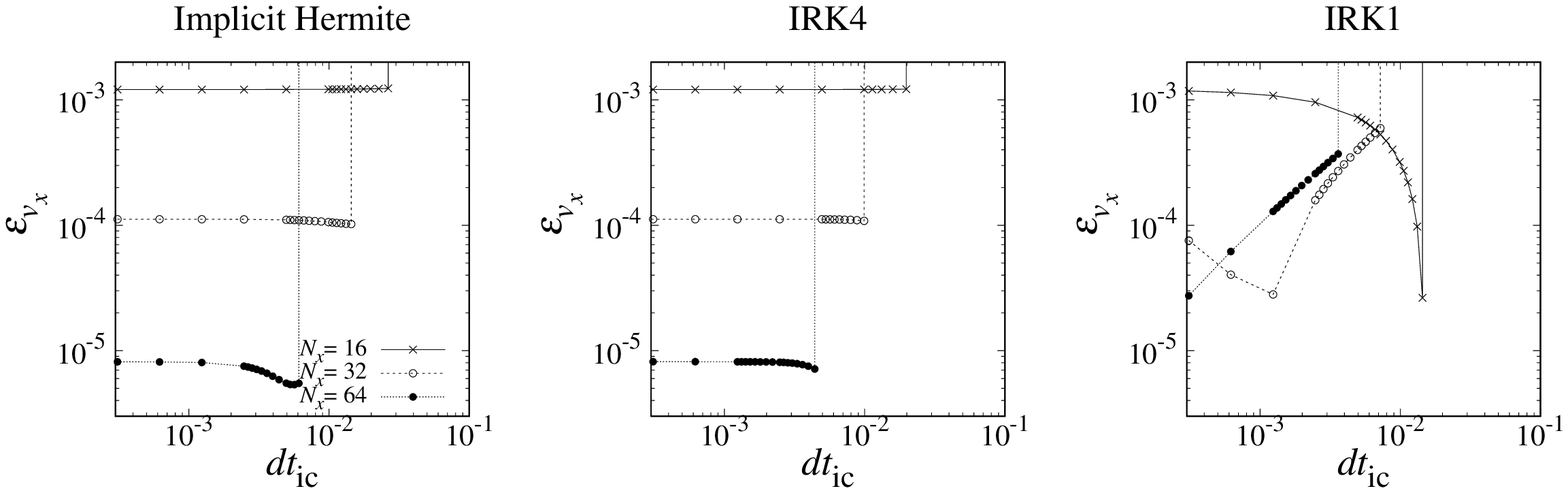} 
	\caption{
			The panel shows $\epsilon_{v_x}$ plotted against
			$dt_{\rm ic}$.
			Left, middle and right sides panels show the results
			for implicit Hermite scheme , IRK4 and IRK1.
			Crosses, open and filled circles show results
			for $N_x=16$, $32$ and $64$.
		}
		\label{fig:gravwaveerr}
	\end{center}
 \end{figure}

Figure \ref{fig:gravwaveerr} shows errors plotted against $dt_{\mathrm{ic}}$.
The result shows that the accuracy of the implicit Hermite scheme is
similar to that of IRK4 and smaller than that of IRK1 with large $N$.

\begin{figure}[ht]
	\begin{center}
	\includegraphics[width=7cm,clip]{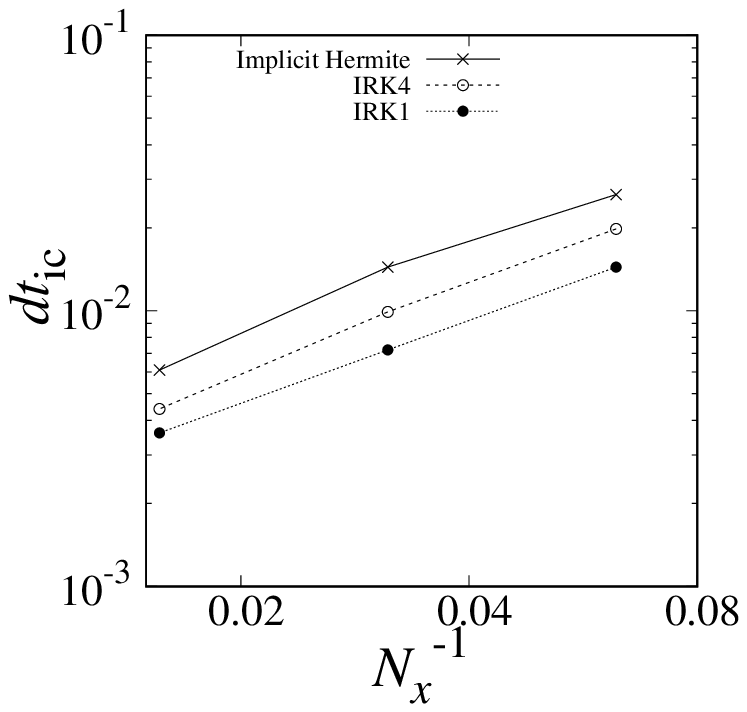} 
	\caption{
			The panels show maximum $dt_{\mathrm{ic}}$ in the numerical
			stable region for tests with $x_0 = 0.006$ plotted
			against $N_x^{-1}$.
			Crosses, open and filled circles show the results of the
			implicit Hermite scheme, IRK4 and IRK1.
		}
		\label{fig:gravwaveLimitDt}
	\end{center}
 \end{figure}

Figure \ref{fig:gravwaveLimitDt} shows the maximum $dt_{\mathrm{ic}}$ in the
numerical stable region plotted against $N_x^{-1}$.
We can see that the region of stability of the implicit Hermite
scheme are wider than those of IRK1 and IRK4.
Hence, we can use larger timesteps with the implicit Hermite scheme.
Therefore, we can conclude that the Hermite scheme is better than
Runge-Kutta schemes for simulations of fluid with the surface and gravity
wave.

\section{Summary}
\label{sec:discandsum}
If we use multi-stage integration schemes, such as Runge-Kutta schemes, with
mesh-free methods we need to perform the interaction calculation, which
is the most expensive part of the calculation, multiple times per one time
step. We constructed the Hermite scheme for a high-order
mesh-free method. 
The accuracy of fourth-order Hermite schemes is 
at least similar to those of Runge-Kutta schemes and
the region of stability of Hermite schemes are better than those of
Runge-Kutta schemes. Therefore, we can use a large time step with the Hermite
scheme compare to that for the Runge-Kutta scheme for the same accuracy.
We conclude that Hermite schemes are more computationally efficient
than commonly used Runge-Kutta schemes for a high-order mesh-free method.


\begin{ack}
	We would like to thank the referee for his or her insightful comments
	and suggestions.
	We also thank the editor for his or her assistance.
	We thank Masaki Iwasawa, Keigo Nitadori and Daisuke Namekata for
	discussions about Hermite schemes and Runge-Kutta schemes.
	This research was supported by RIKEN Junior Research Associate
	Program and MEXT as ``Exploratory Challenge on
	Post-K computer'' (Elucidation of the Birth of Exoplanets [Second
	Earth] and the Environmental Variations of Planets in the Solar
	System).
\end{ack}





\begin{thebibliography}{}
%
%
\bibitem[Antuono et al.(2011)]{2011CoPhC.182..866A}
	Antuono, M., Colagrossi, A., Marrone, S., \& Lugni, C.\ 2011, Computer Physics Communications, 182, 866 
\bibitem[Aoki(1997)]{Aoki1997IDO}
	Aoki, T.\ 1997, Computer Physics Communications, 102, 132
\bibitem[Gingold \& Monaghan(1977)]{1977MNRAS.181..375G}
	Gingold, R.~A., \& Monaghan, J.~J.\ 1977, \mnras, 181, 375
\bibitem[Makino(1991)]{1991ApJ...369..200M}
	Makino, J.\ 1991, \apj, 369, 200
\bibitem[Makino \& Aarseth(1992)]{1992PASJ...44..141M}
	Makino, J., \& Aarseth, S.~J.\ 1992, \pasj, 44, 141
\bibitem[Sod(1978)]{sod1978survey}
	Sod, G.A.\ 1978, Journal of computational physics, 27, 1
\bibitem[Wendland(1995)]{wendland1995piecewise}
	Wendland, H.\ 1995, Advances in computational Mathematics, 4, 389.
\bibitem[Yamamoto \& Makino(2017)]{2017arXiv170105316Y}
	Yamamoto, S., \& Makino, J.\ 2017, \pasj, 69.
\end{thebibliography}
\end{document}